\documentclass[fleqn,usenatbib]{mnras}

\usepackage{newtxtext,newtxmath}
\usepackage{xcolor}
\usepackage{siunitx}

\usepackage[T1]{fontenc}

\DeclareRobustCommand{\VAN}[3]{#2}
\let\VANthebibliography\thebibliography
\def\thebibliography{\DeclareRobustCommand{\VAN}[3]{##3}\VANthebibliography}

\usepackage{graphicx}	
\usepackage{amsmath}	

\title[Characterization of kilonova emission]{Impact of jets on kilonova photometric and polarimetric emission from binary neutron star mergers}

\author[M. Shrestha et al.]{
Manisha Shrestha,$^{1,2}$\thanks{E-mail: mshrestha1@arizona.edu}
Mattia Bulla,$^{3,4,5,6}$
Lorenzo Nativi,$^{6}$
Ivan Markin,$^{7,8}$
Stephan Rosswog,$^{6,9}$ 
\newauthor
Tim Dietrich$^{7,10}$
\\
$^{1}$Steward Observatory, University of Arizona, 933 North Cherry Avenue, Tucson, AZ 85721-0065, USA\\
$^{2}$Astrophysics Research Institute, Liverpool John Moores University, Liverpool Science Park IC2, 146 Brownlow Hill, UK \\
$^{3}$Department of Physics and Earth Science, University of Ferrara, via Saragat 1, I-44122 Ferrara, Italy\\
$^{4}$INFN, Sezione di Ferrara, via Saragat 1, I-44122 Ferrara, Italy\\
$^{5}$INAF, Osservatorio Astronomico d’Abruzzo, via Mentore Maggini snc, 64100 Teramo, Italy\\
$^{6}$Department of Astronomy and Oskar Klein Centre, Stockholm University, AlbaNova, SE-10691 Stockholm, Sweden \\
$^{7}$Institute for Physics and Astronomy, University of Potsdam, Haus 28, Karl-Liebknecht-Str. 24/25, 14476 Potsdam, Germany \\
$^{8}$Leibniz-Institute for Astrophysics Potsdam (AIP), An der Sternwarte 16, 14482 Potsdam, Germany\\
$^{9}$Hamburger Sternwarte, University of Hamburg, Gojenbergsweg 112, 21029 Hamburg, Germany \\
$^{10}$Max Planck Institute for Gravitational Physics (Albert Einstein Institute), Am M\"{u}hlenberg 1, 14476 Potsdam, Germany\\
}

\date{Accepted XXX. Received YYY; in original form ZZZ}

\pubyear{2022}

\begin{document}
\label{firstpage}
\pagerange{\pageref{firstpage}--\pageref{lastpage}}
\maketitle

\begin{abstract}

A merger of binary neutron stars creates heavy unstable elements whose radioactive decay produces a thermal emission known as a kilonova. In this paper, we predict the photometric and polarimetric behaviour of this emission by performing 3-D Monte Carlo radiative transfer simulations. In particular, we choose three hydrodynamical models for merger ejecta, two including jets with different luminosities and one without a jet structure, to help decipher the impact of jets on the light curve and polarimetric behaviour. In terms of photometry, we find distinct color evolutions across the three models. Models without a jet show the highest variation in light curves for different viewing angles. In contrast, to previous studies, we find models with a jet to produce fainter kilonovae when viewed from orientations close to the jet axis, compared to a model without a jet. In terms of polarimetry, we predict relatively low levels ($\lesssim0.3-0.4$\%) at all orientations that, however, remain non-negligible until a few days after the merger and longer than previously found. Despite the low levels, we find that the presence of a jet enhances the degree of polarization at wavelengths ranging from $0.25$ to $2.5\micron$, an effect that is found to increase with the jet luminosity. Thus, future photometric and polarimetric campaigns should observe kilonovae in blue and red filters for a few days after the merger to help constrain the properties of the ejecta (e.g. composition) and jet.

\end{abstract}




\section{Introduction}
The merger of compact objects such as binary neutron star (NS) or black hole (BH) - NS systems can produce gravitational wave (GW) signals along with electromagnetic (EM) counterparts. This was confirmed by the 17th August 2017 detection of a short gamma-ray burst (GRB) 170817A \citep{Goldstein_2017,Savchenko_2017} followed by the kilonova (KN) emission AT2017gfo \citep{Coulte_2017} in coincidence with the GW event GW170817 \citep{Abbott_2017} from the merger of a binary NS system. Due to a good localization of the event, many ground based telescopes could follow-up this watershed event throughout the entire electromagnetic spectrum \citep[e.g.][]{ Alexander_2017,Andreoni_2017, Arcavi_2017, Covino_2017, Cowperthwaite_2017, Drout_2017, Evans_2017, Haggard_2017, Kasliwal_2017, Margutti_2017, Pian_2017, Smartt_2017, Soares-Santos_2017, Tanvir_2017, Troja_2017, Utsumi_2017,Valenti_2017}. This event opened the new era of multi-messenger astrophysics with gravitational wave sources.

KN emission is produced by the radioactive decay of the unstable heavy elements produced during the merger event and it was first predicted  based on a simple, semi-analytical model by \cite{Li_1998}. The optical emission from the KN can last for a few days, thus making it complementary to observations of short-lived GRB. These mergers are major sources of  r-process elements \citep{lattimer74,symbalisty82,eichler89,rosswog99,freiburghaus99} 
and the neutron-richness of the ejecta makes them in particular excellent candidates for the 3rd r-process peak
containing e.g. gold and platinum.
Various groups identified the r-process elements in  AT2017gfo \citep{Watson_2019,Domoto_2021,Kasliwal2022} which points to either all or part of them being formed in binary NS mergers. 

Even though these mergers are considered primary sources of heavy elements, there is only one confirmed KN observation in the form of AT2017gfo. There are a few possible candidates for KN such as KN associated with GRB 130603B \citep{Tanvir_2013} and GRB 211211A \citep{Rastinejad_2022, Troja_2022}. There are many open questions such as the density distribution of merger ejecta, distribution of lanthanide-rich material in the ejecta, and velocity of the ejecta. A combination of polarimetric and photometric studies will be crucial in improving our understanding of these open questions.
 Due to the lack of a wide array of observed KN, we need to rely on simulations to predict the properties of future KN observations as well as how to better equip various telescopes for efficient observations of these entities. There are various simulations that are designed for predicting spectra and light curves of KN emission \citep[e.g.][]{Wollaeger_2018,Kawaguchi2018,Bulla_2019b,Darbha2020,Korobkin2021,Collins2022} and few include polarization signal predictions \citep{Bulla_2019, Bulla_2021, Matsumoto_2018, Li_2019}. We build on the work by \citet{Nativi_2021} and \citet{Klion_2021} where the impact of the jet on the ejected material and corresponding kilonova emission was investigated. In this paper, we add on to the models to make them more realistic by including a dynamical ejecta component and time-evolving opacities. Finally, we present both polarimetric and photometric results. We make use of the 3-D Monte Carlo radiative transfer (MCRT) code \textsc{possis} for the simulations \citep{Bulla_2019b,possis2}. 
 
 The paper is organized as follows: in Section~\ref{sec:Methods}, we present the general method used for our simulation together with the simulation setup of specific models. Then, we present results in Section~\ref{sec:results}, which is divided into photometric results (Section~\ref{subsec:Photometry}) and polarimetric results (Section~\ref{subsec:Polarimetry}). In Section~\ref{sec:Discussions}, we discuss the implications of these results and how these results could be useful for future observation planning and analysis. Finally, we provide concluding remarks in Section~\ref{sec:Conclusions}.


\section{Methods} \label{sec:Methods}

\begin{figure*}
    \centering
    \includegraphics[scale=0.42]{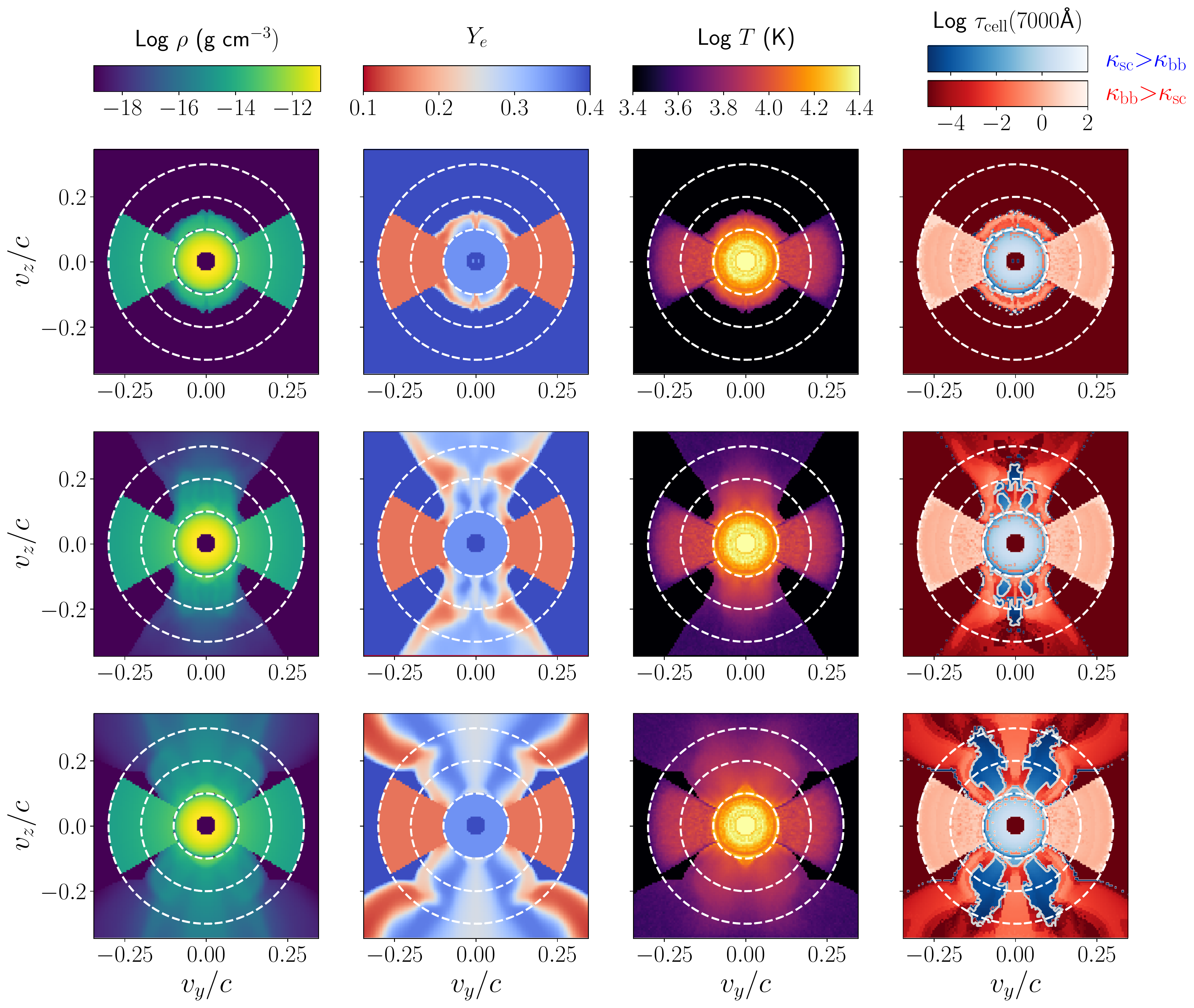}
    \caption{Model properties at 1 day after merger for \texttt{Wind} (top), \texttt{Jet49} (middle), and \texttt{Jet51} (bottom). From left to right, maps in the velocity $v_y-v_z$  plane are shown for the density $\rho$, the electron fraction $Y_e$, the temperature $T$, and the optical depth at $7000\AA$ within each cell $\tau_{\rm cell}=(\kappa_{\rm es}+\kappa_{\rm bb})\,\rho\,dr$, where $\kappa_{\rm es}$ and $\kappa_{\rm bb}$ are the Thomson scattering and bound-bound opacities and $dr$ is the cell width. The optical depths are shown in red (blue) colors for cells where bound-bound opacity is larger (smaller) than electron-scattering opacity. Light grey contours in the right panels highlight the electron-scattering plumes where electron scattering opacity is dominant over bound-bound opacity.  Dashed white circles mark velocities of 0.1, 0.2, and 0.3c. All the models have a $128^3$ grid resolution. 
    }
    \label{fig:modelprop-0.5}
\end{figure*}

The simulations in this work are carried out using the latest \citep{possis2} version of \textsc{possis} \citep{Bulla_2019b}, a 3-D MCRT code that has been used in the past to predict polarization signatures of astrophysical transients as supernovae \citep{Bulla_2015,Inserra_2016}, kilonovae \citep{Bulla_2019,Bulla_2021} and tidal disruption events \citep{Leloudas2022,Charalampopoulos2022}. The code simulates the propagation of $N_{\rm ph}$ Monte Carlo (MC) photon packets throughout a medium expanding homologously and calculates flux and polarization spectra as a function of time $t$ and observer viewing angle $\theta_{\rm obs}$. The energy of each MC packet is initialized by splitting into equal parts \citep{Abbott_1985,Lucy_1999} the total energy available from the radioactive decay of $r-$process nuclei, with heating rates taken from \citet[][see their equation 2]{Rosswog2022}  and thermalization efficiencies computed following \cite{Barnes2016} and \cite{Wollaeger_2018}. Each MC packet is assigned a normalized Stokes vector \textbf{s} = (1,$q$,$u$) that is initialized to \textbf{s$_0$} = (1,0,0), i.e. packets are created with no polarization. The propagation of MC packets is controlled by the opacity of the expanding material. Time-dependent, state-of-the-art opacities from \cite{Tanaka_2020} are adopted for both bound-bound line transitions, $\kappa_{\rm bb}(\lambda,t,\rho,T,Y_e)$, and electron scattering, $\kappa_{\rm es}(t,\rho,T,Y_e)$, as a function of local properties of the ejecta such as density $\rho$, temperature $T$ and electron fraction $Y_e$. MC packets are polarized by electron scattering and depolarized by bound-bound transitions, with their Stokes vectors updated following \cite{Bulla_2015}.  Spectra are extracted using ``virtual'' packets as described in \cite{Bulla_2015}, with this technique reducing significantly the MC
noise in the spectra compared to the angular binning of escaping packets more commonly adopted in the literature. 
We refer the reader to \cite{Bulla_2015}, \cite{Bulla_2019b}, and \cite{possis2} for more details about the code.

Polarization spectra are computed for three different models from \cite{Nativi_2021}. In these models, a neutrino-driven wind from \cite{Perego_2014} has been evolved assuming that no jet (model \texttt{Wind}) or jets with opening angle $\theta_{0}=5^\circ$ and luminosity of $L_{\rm j}=10^{49}$\,erg\,s$^{-1}$ (model \texttt{Jet49}) and $L_{\rm j}=10^{51}$\,erg\,s$^{-1}$ ( model \texttt{Jet51}) have been launched and propagated through the ejecta for 100 ms. In addition, all models include a spherical component at low velocities ($0.03-0.1$c, where c is the speed of light) accounting for ``secular'' ejection mechanisms from the disk torus around the merger remnant \citep[e.g.][]{beloborodov08,Just_2015,Siegel_2018,fernandez_2019,Miller_2019}. This component is relatively massive ($M_{\rm sec}=0.072\,M_\odot$) and in this work assumed to have a fixed $Y_e=0.35$, although the detailed composition of the secular ejecta is still a matter of debate \citep[e.g.][]{Siegel_2018,Miller_2019}. The main focus of this work is the polarimetric behaviour of KN emission, hence we do not investigate the lanthanide-rich composition explored in \cite{Nativi_2021} as it would lead to a null polarization signal due to electron scattering being subdominant (see discussion in \citealt{Bulla_2019}).
Compared to \cite{Nativi_2021}, here we model an additional component to account for material ejected dynamically. Specifically, we assume a dynamical-ejecta component with a mass $M_{\rm dyn}=0.005\,M_\odot$, a lanthanide-rich composition ($Y_e=0.15$) and a distribution extending from 0.1 to 0.3c and in a conical region around the merger plane with half-opening angle $\phi=30^\circ$. 

Density, $Y_e$, temperature, and opacity maps at $1$\, day after the merger are shown in Figure~\ref{fig:modelprop-0.5} for the three different models. We outline regions in \texttt{Jet49} and \texttt{Jet51} important for interpreting the polarization signals --- the regions of the wind material propelled by the jet at $\theta \sim \ang{25}$ from the polar axis, with high $Y_e$ and dominated by electron-scattering (light grey contours in the right panels of Fig.~\ref{fig:modelprop-0.5}). These regions are more extended in the \texttt{Jet51} compared to \texttt{Jet49} model. Hereafter, we refer to these regions as electron-scattering plumes.


Radiative transfer simulations presented in this work are carried out for $N_{\rm ph}=5\times10^8$ MC photons and for a grid resolution of $128^3$ for all three models. Flux and polarization spectra are extracted for $100$ time bins logarithmically spaced from $0.1$ and $30$ days after the merger, $1000$ wavelength bins logarithmically spaced from $0.05$ and $10\,\mu$m and $11$ viewing angles equally spaced in cosine from $\cos\theta_{\rm obs}=1$ (face-on, along the jet axis) to $\cos\theta_{\rm obs}=0$ (edge-on, in the merger plane). Since the models are axially symmetric around the jet axis, the polarization signal is carried by Stokes $q$ while Stokes $u$ is consistent with zero, and its signals are used as a proxy for MC noise. 

The modeling performed here is an improvement over previous studies in various respects. First, we take into account time-dependent effects that were not considered in polarization studies of \cite{Bulla_2019} and \cite{Bulla_2021}, in which ejecta properties (including opacities) were frozen at selected epochs. Secondly, we model all the main components expected in binary neutrons star mergers -- jet, wind, dynamical and secular ejecta -- instead of restricting to a broad two-component model as done e.g. in \cite{Bulla_2019}. Finally, we adopt state-of-the-art opacities as a function of local properties of the ejecta (see above) in place of uniform values in each component as done in both \cite{Bulla_2019} and \cite{Bulla_2021} or approximate analytic functions as done in \cite{Nativi_2021}. 

\section{Results} \label{sec:results}

\begin{figure*}
    \centering
    \includegraphics[scale=0.425]{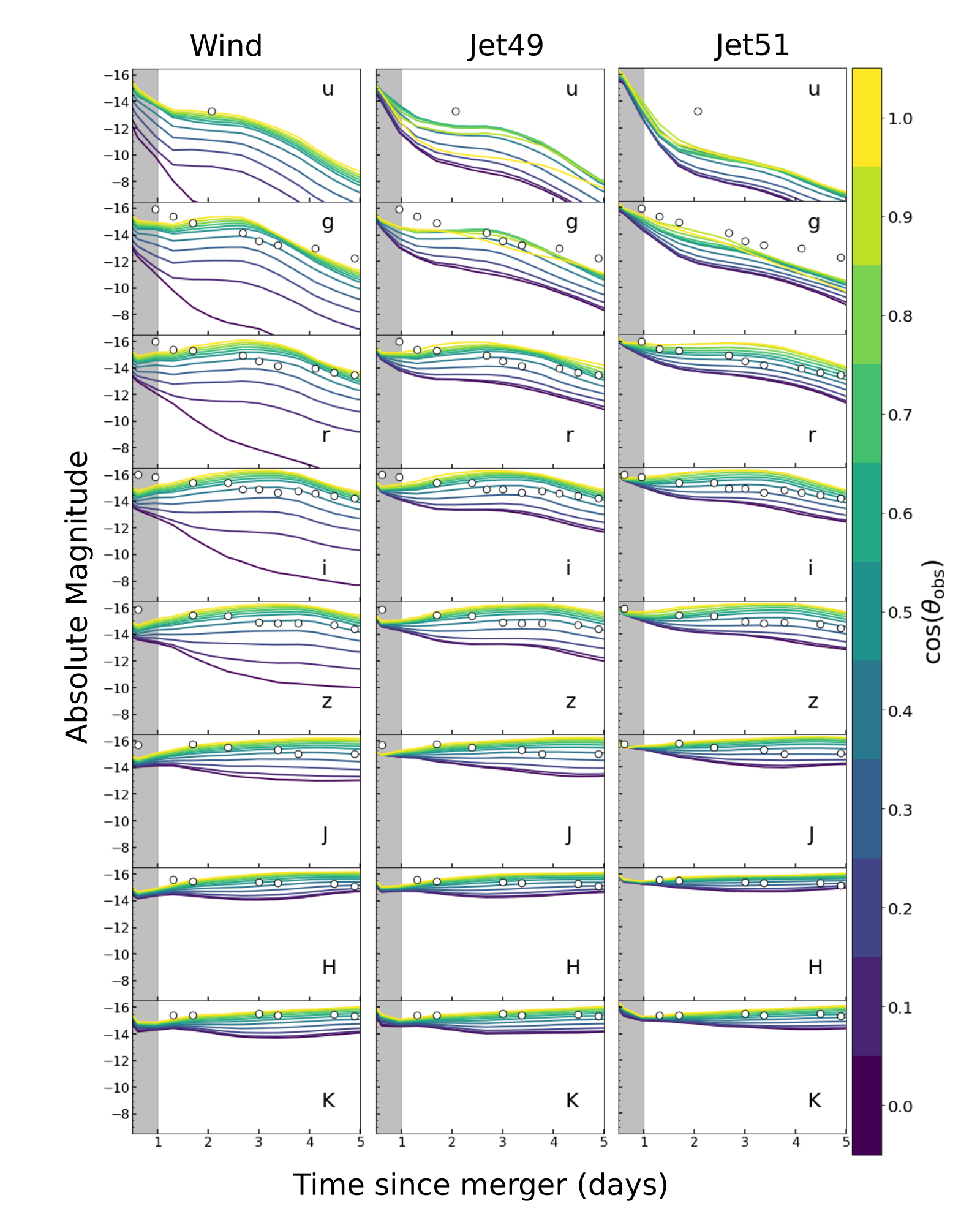}
    \caption{Broadband $ugrizJHK$ light curves of three models \texttt{Wind} (left), \texttt{Jet49} (middle), and \texttt{Jet51} (right). Different colored lines represent the light curve at various inclination angles. The open circle is the observational data of AT2017gfo which has been corrected for the Milky-Way extinction using \citet{Schlafly_2011}.  The shaded gray region represents the time period up to 1.0 days after the merger where opacity calculations are not highly reliable \citep{Tanaka_2020}. }
    \label{fig:lc}
\end{figure*}

\begin{figure*}
    \centering
    \includegraphics[scale=0.25]{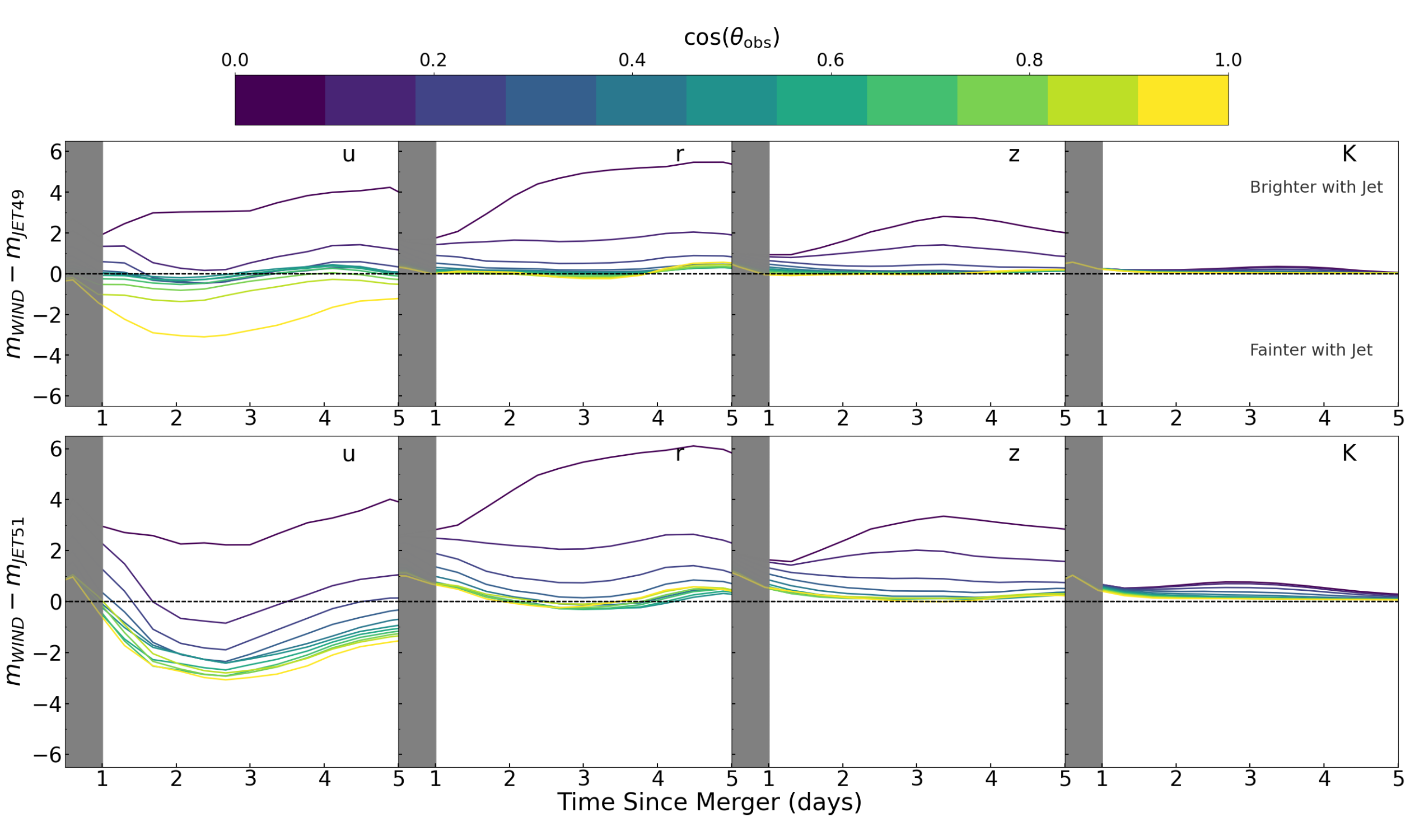}
    \caption{Difference in magnitude for different filters for the jet models compared to wind model. The top plot is for the difference in \texttt{Wind} and \texttt{Jet49} magnitude and the bottom is for the difference in \texttt{Wind} and \texttt{Jet51} magnitude. The eight different panels are for different filters $urzK$ respectively. As in Fig.~\ref{fig:lc}, different colored lines in each panel represent different inclination angles ranging from $0$ to $90$ degrees from dark blue to yellow respectively. We have labeled the regions below and above the zero as the case where \texttt{Jet} models are fainter and brighter than the \texttt{Wind} model which is true for both panels.} 
    \label{fig:magdiff}
\end{figure*}

In this section, we present the results from the three models described in Section~\ref{sec:Methods}. First, we present the photometric results in Section~\ref{subsec:Photometry} and then provide the polarimetric results in the form of spectropolarimetry and polarimetric curve in Sections~\ref{subsubsec:polspec} and \ref{subsubsec:polcurve}, respectively. 

\subsection{Photometry}\label{subsec:Photometry}

From the spectral time series obtained with \textsc{possis}, we can construct light curves for different filters and different viewing angles. In Fig.~\ref{fig:lc}, we present the results for three models: \texttt{Wind}, \texttt{Jet49}, and \texttt{Jet51}. Different panels in Fig.~\ref{fig:lc} refer to the different filters $u,g, r, i, z, J, H,$ and $K$. Each panel has light curves for different inclination angles going from $0\degr$ to $90 \degr$ represented by dark blue to yellow colors respectively. We have included observed data of AT2017gfo as open circles in our light curve to present comparative values to the readers. However, we are not doing a rigorous comparison between our model and AT2017gfo data, and there is no reason to expect that the employed models have to describe the AT2017gfo observation.


The time evolutions of the light curves of \texttt{Jet49} and \texttt{Jet51} show a similar trend to the \texttt{Wind} model. However, the variation in apparent magnitude with respect to the inclination angle is smaller in the models with a jet compared to models without a jet, an effect that was also seen by \citet{Nativi_2021}. This behaviour can be understood based on the opacity distribution of these models as shown in Fig.~\ref{fig:modelprop-0.5}. As shown in the opacity maps, the presence of a jet distributes some optically thick material from the wind component to higher velocities ($\gtrsim0.1$c). Therefore, some of the photons emitted towards the polar regions can get scattered/reprocessed to other angles. In addition, regions of the wind which in the \texttt{Wind} model were obscured by the dynamical ejecta when viewed from equatorial viewing angles are now exposed to these orientations in the \texttt{Jet49} and \texttt{Jet51} models. Compared to the wind case, the kilonova with the jet cases is, therefore, fainter at polar angles and brighter at equatorial/intermediate angles, effectively reducing the viewing angle dependence. This effect can also be seen in the magnitude differences presented in Fig.~\ref{fig:magdiff}. These predictions are in contrast to those by \citet{Nativi_2021} and \citet{Klion_2021} and will be further discussed in Section~\ref{sec:Discussions}. 

The evolution of light curves in different filters for all the models is similar in nature. We see that for shorter wavelengths $u$, and $g$, the brightness decreases with time for all the inclinations angles. However, the sharpness of the decrease in brightness is dependent on the type of model. For the case of \texttt{Wind}, the decrease in brightness is gradual, for \texttt{Jet49} the decrease is sharper and the decrease is the sharpest for \texttt{Jet51}. This can be explained by the variation in scattering opacity along the jet axis based on wavelengths with time. The opacity is higher in shorter wavelengths ($u$, and $g$) and it increases with time. Thus, more photons with shorter wavelengths are scattered. Therefore, only a small portion of photons in this wavelength regime can escape for the models with jet, which produces the rapid decline in brightness  in $u,g$ filters. In $r,i,z, J, H, K$ filters, there is an initial decline in brightness followed by an increase and then a plateau for viewing angles closer to the equatorial plane. For polar viewing angles, the brightness increases and then plateaus. Both equatorial and polar behaviour can be explained in terms of optical depth evolution. As time evolves, the optical depth decreases, thus the multiply interacting photons from earlier times are able to escape from the simulation grid. As photons are reprocessed by lines, they are re-emitted at longer wavelengths. Hence, the kilonova signal is higher at these longer wavelengths.



\subsection{Polarimetry} \label{subsec:Polarimetry}
 One of the main aims of the paper is to study the impact of a jet on polarization signal from KNe emission. From the simulations described in Section~\ref{sec:Methods}, we get polarimetric information in the form of Stokes vectors $q$ and $u$. Since the models are symmetric around the jet axis, Stokes $u$ is expected to be zero, and Stokes $q$ quantifies the polarization signal. We can create polarization spectra as shown in Fig.~\ref{fig:specpol} and from that, we can create a broadband polarimetric curve as shown in Fig.~\ref{fig:polcurve}. In this section, we present the polarimetric results for the three different models \texttt{Wind}, \texttt{Jet49}, and \texttt{Jet51}.

\subsubsection{Polarization Spectra} \label{subsubsec:polspec}

\begin{figure*}
    \centering
    \includegraphics[width=\textwidth]{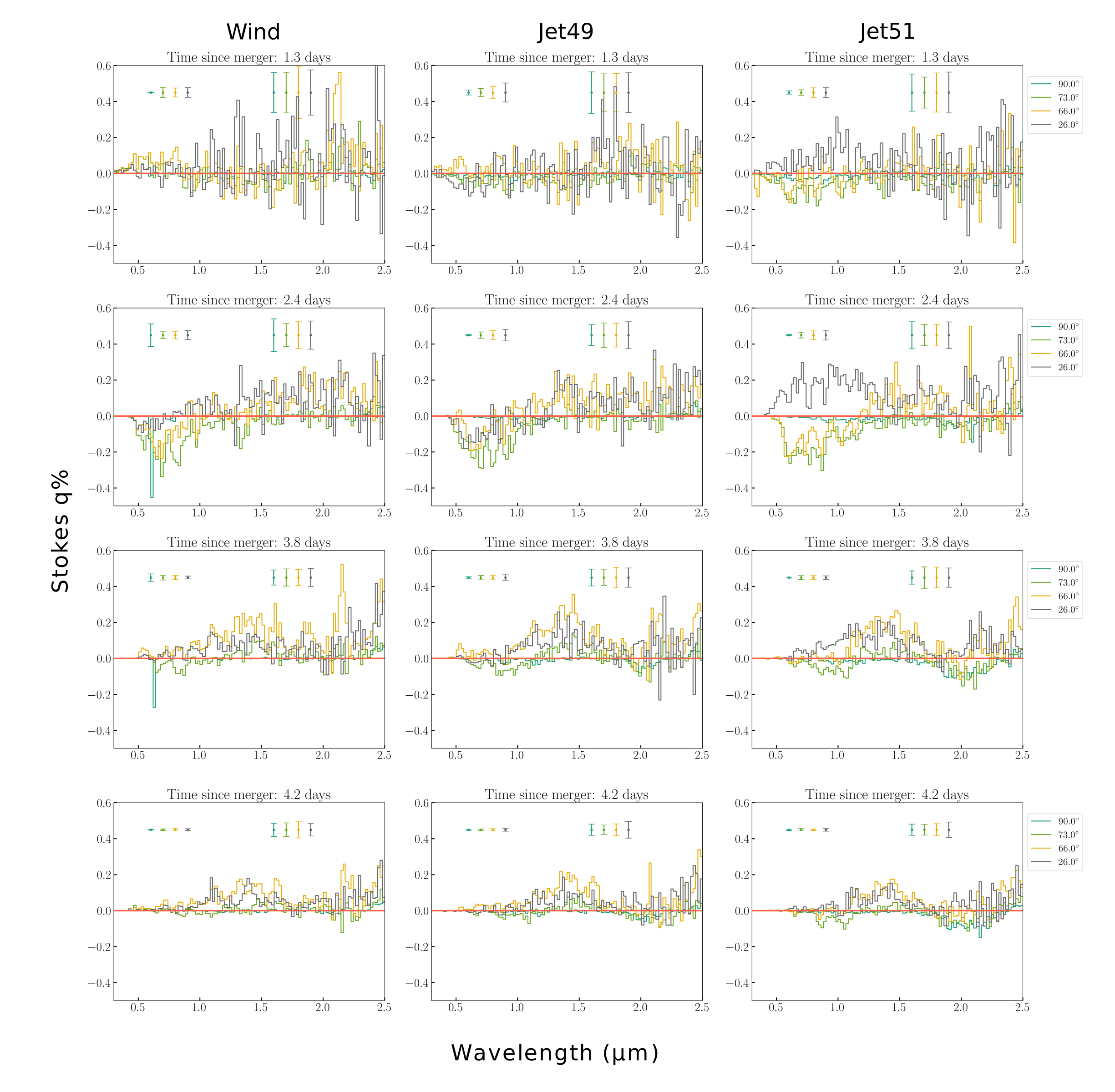}
    \caption{Spectropolarimetric curves for the three different models from left to right. Different panes top to bottom is for different time since the merger. And different color curves in each panel represent different viewing angles. The scatter in Stokes $u$ is assumed to be the error in Stokes $q$. For each viewing angle, we present one error value} for wavelengths shorter than $1.5 \mu$ in the top-left corner of each panel and another error bar for longer wavelengths in the top-right corner. 
    \label{fig:specpol}
\end{figure*}

In Fig.~\ref{fig:specpol}, we present the variation for Stokes $q$ with respect to wavelengths for the three different models \texttt{Wind}, \texttt{Jet49}, and \texttt{Jet51} for four selected time epochs of $1.3$, $2.4$, $3.8$, and $4.2$ days after the merger and four different inclination angles of  $\theta_{\rm obs}=26 \degr, 66 \degr, 73 \degr$ and $90\degr$. We see that polarization is highly dependent on the inclination angle and the evolution with time for the three models is different from each other. 

For the case of \texttt{Wind}, we mostly detect very low levels of Stokes $q$ for all the filters and viewing angles and we detect some higher scatter at day 1.3 after the merger. Since the error bar is larger, we can attribute this mostly to MC noise. The highest level of polarization is seen for the intermediate angle of $\theta_{\rm obs}=73 \degr$, with a low Stokes $q \sim -0.28 \% $ at 2.4 days after the merger. The low level of polarization can be understood from the opacity map shown in Fig.~\ref{fig:modelprop-0.5}. The region with dominant electron scattering opacity, i.e. blue region is associated with the secular component, is mostly concentrated around the core, and is spherically symmetric. The photons that are scattered from this portion have to go through the higher density regions with the possibility of absorption or multiply scattered. This can cause the polarization value to be lower. The small polarization signal in $q$ is comparable to the $u$ signal shown as  error bars in the plot. This shows that the signal in Stokes $q$ is compatible with MC noise.

For the \texttt{Jet49} model, the highest polarization values are seen for $\theta_{\rm obs}=26$ and $66 \degr$ at wavelengths shorter than $1.0 \micron$ in earlier time periods. At $2.4$ days after the merger, a polarization signal of negative $0.3\%$ is seen which decreases on $3.8$ days after the merger with a polarization signal closer to negative $0.25\%$. On days 3.8 and 4.2 after the merger, we see some positive polarization signal at wavelengths longer than $1.25 \mu m$ as shown in Fig.~\ref{fig:specpol}. This could be attributed to photons that emerge from lanthanide-rich dynamical ejecta at late times and infrared wavelengths and are scattered by electron-scattering plumes before reaching the observer.
Initially, the density of dynamical ejecta in the equatorial plane is extremely high, thus the photons scattered in dynamical ejecta cannot escape the simulation grid in earlier days like 1.3 and 2.4 days after the merger. However, the electron scattering opacity is higher in regions close to the jet axis as shown by the blue regions or the electron-scattering plumes in Fig.~\ref{fig:modelprop-0.5}. This can scatter photons and produce a polarized signal. Since the photons are getting scattered from polar regions, the scattered photons have preferentially negative Stokes $q$ which can be seen in Fig.~\ref{fig:polmaps}. These are a few photons that can escape the grid in earlier times with less scattering, hence this negative Stokes $q$ is seen for shorter wavelength. As time increases, the densities in the dynamical ejecta decrease and the photons that are scattered from the electron-scattering plume region can escape with a positive sign in their Stokes $q$.

\begin{figure*}
    \centering
    \includegraphics[scale=0.325]{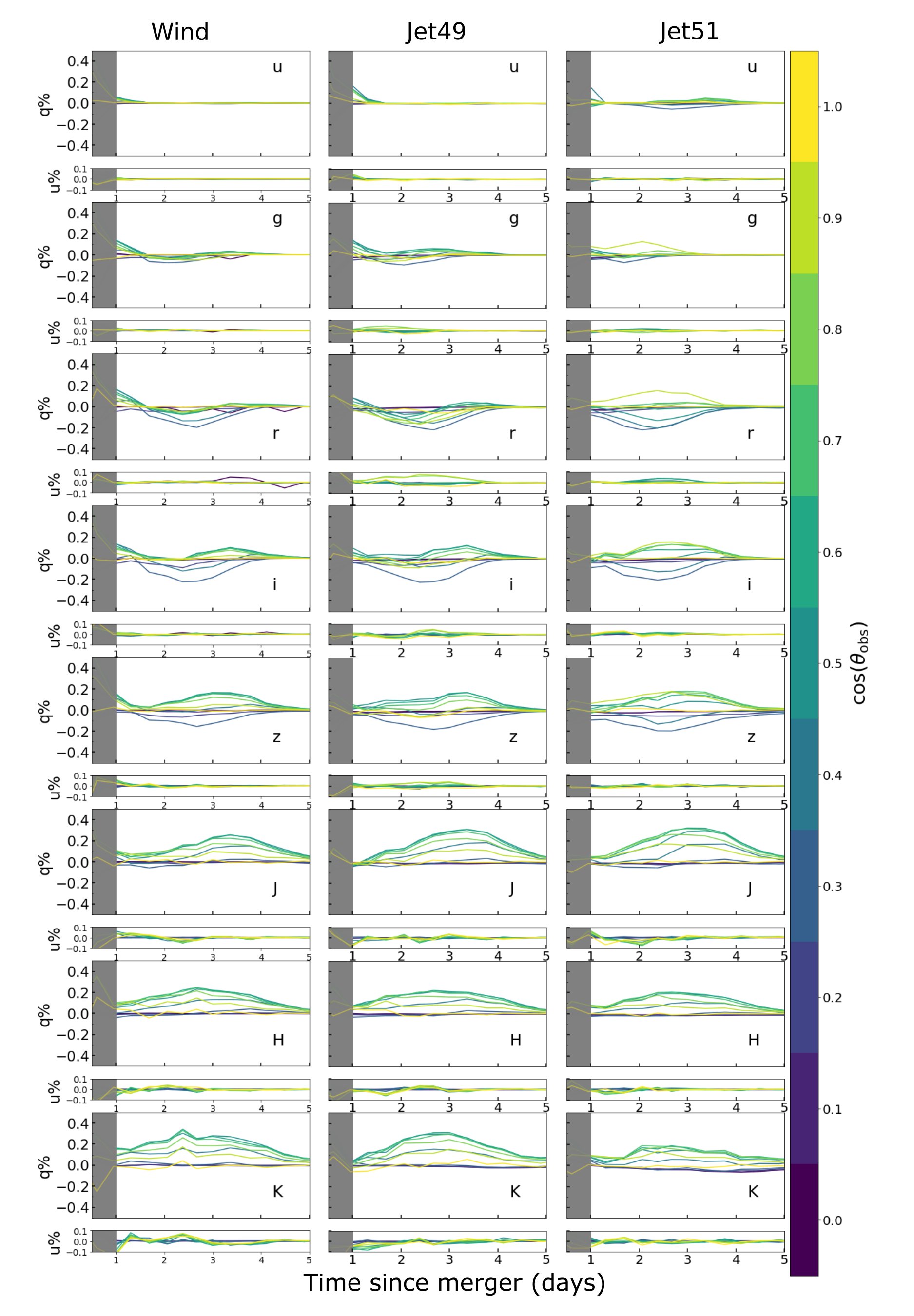}
    \caption{Plot of $\%q$ and $\%u$ with respect to time for $ugriz$ broadband filters. Three columns of plots are for three different models. The color of different curves represents the different inclination angles. The shaded gray region represents the time period of 0.5 to 1.0 days after the merger where opacity calculations are not highly reliable. }
    \label{fig:polcurve}
\end{figure*}

The \texttt{Jet51} model shows higher levels of polarization for all the wavelengths on 2.4 days after the merger compared to \texttt{Wind} and \texttt{Jet49} case. This directly relates to a larger region of electron-scattering plumes. At 1.3 days after the merger, there is no clear polarization signal. However, we predict negative Stokes $q$ for $\theta_{\rm obs}=66$ and $73 \degr$ and for wavelength less than $0.85 \mu m$ which can also be seen in Fig.~\ref{fig:polmaps} for the epoch of 2.4 days after the merger. For $\theta_{\rm obs}=26 \degr$  there is some positive Stokes $q$ signal. For the other epochs, the polarization signal for this model is similar to cases of \texttt{Wind} and \texttt{Jet49}.

\subsubsection{Polarization curves}\label{subsubsec:polcurve}

The results from Section~\ref{subsubsec:polspec} can be converted to polarization curves for different broadband filters ($ugriz$)  as shown in Fig.~\ref{fig:polcurve}. 
Due to the wider availability of imaging polarimeters compared to spectropolarimeters and imaging polarimetry requiring fewer photons for better polarization accuracy, polarization curve predictions can be crucial from an observational standpoint.
Hence, in this section, we present the prediction of polarization evolution with time for various optical and NIR filters. Different colors in each panel are for different inclination angles. 

The \texttt{Wind} model's polarization curve in Fig.~\ref{fig:polcurve} (left panels) shows mostly a low level of polarization signal. We see a slight increase in polarization at later times for longer wavelengths like $rizJHK$ filters and the noise level is low as well indicating it to be an actual polarization signal. For viewing angles closer to the edge-on, we detect some negative Stokes $q$ which is more prominent in $riz$ filters up to 3 days after the merger with the peak at 2.5 days after the merger with values between $0.2 -  0.25 \%$. After 2.5 days we detect positive Stokes $q$ ($\lesssim 0.3\%$) for viewing angle towards to the polar regions for $JHK$ filters. Higher polarization in longer wavelengths at later times can be attributed to multiply-interacting photons escaping at later times from higher opacity dynamical ejecta regions. The polarization signal about the inclination angle of $0\degr$ is very close to zero due to the symmetry of the geometry. Hence the signal we are getting of a few percent is real and not MC noise. 

The polarization curves of \texttt{Jet49} are presented in Fig.~\ref{fig:polcurve} (middle panels). The polarization curve for $ug$ filters shows an insignificant polarization degree. We see a low level of polarization in $r,i,z,J,H,K$ filters ($q \lesssim 0.3\%$). For $r, i$, and $z$ filters, we see negative Stokes $q$ for days 1 to 3 after the merger. Unlike \texttt{Wind} model, here the negative Stokes $q$ is also seen for viewing angles closer to polar viewing angle. We note that for $riz$ filters we see a significant level of noise during this period, thus we contribute this signal to MC noise. For $JHK$ filters, the positive Stokes $q$ is not accompanied by high noise, thus we believe these signals to be real.  

Fig.~\ref{fig:polcurve} (right panels) shows the results for simulation of \texttt{Jet51} model. In general, we see a slightly higher level of polarization signal for this model compared to the previous two models. For $u$ and $g$ filters, we do not see any clear polarization detection. However, for $rizJHK$ filters, we see mostly positive Stokes $q$ lower than $0.35 \%$ for the viewing angles closer to the polar region. Whereas for more edge-on viewing angles, we detect some negative Stokes $q$. The polarization peak is between 2 and 3 days since the merger for the $riz$ filter. For longer wavelengths $JHK$, there is no clear peak and we see a constant polarization signal of $0.2\%$ from day 1 to day 5 after the merger.

\begin{figure*}
    \centering
    \includegraphics[scale=0.47]{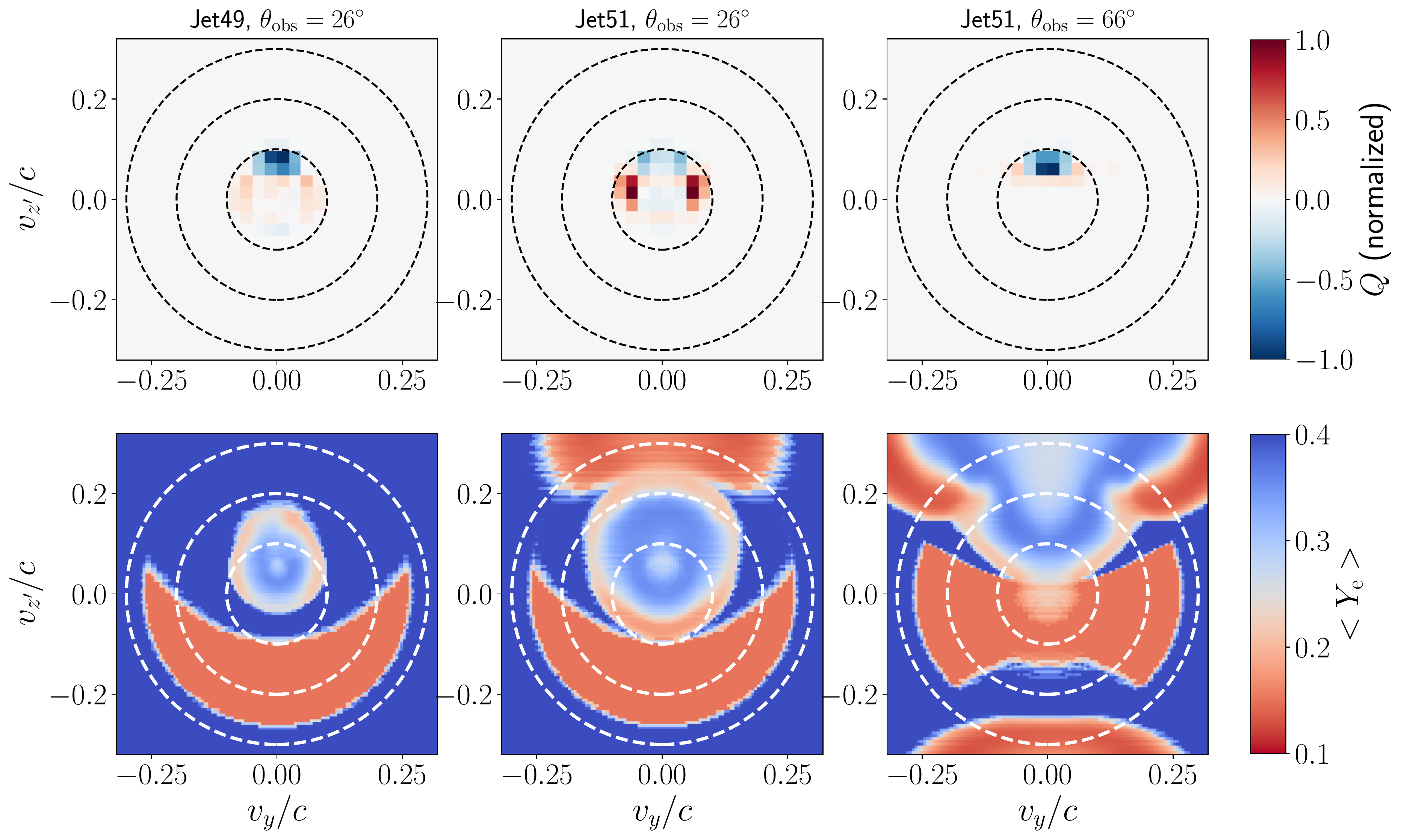}
    \caption{$Q$ maps (top) vs $Y_{\rm e}$ maps (bottom) for three cases from Fig.~\ref{fig:specpol} showing nonzero polarization levels around $2$\,d after the merger and at $\sim7000$\,\AA: the \texttt{Jet49} model viewed from $26\degr$ with $q<0$ (left), the \texttt{Jet51} model viewed from $26\degr$ with $q>0$ (middle) and the \texttt{Jet51} viewed from $66\degr$ with $q<0$ (right). All maps are calculated by rotating the $xyz$ Cartesian grid about the y axis and so that $x'$ corresponds to the line-of-sight to the given observer, i.e. $x'=x\,\cos(\pi/2-\theta_{\rm obs})+z\,\sin(\pi/2-\theta_{\rm obs})$, $y'=y$, $z'=-x\,\sin(\pi/2-\theta_{\rm obs})+z\,\cos(\pi/2-\theta_{\rm obs})$. The $Q$ maps are computed by integrating all the contributions along the line-of-sight and are normalized to the maximum value across the three cases. The $Y_{\rm e}$ maps are computed by averaging the $Y_{\rm e}$ values from $0.075$ to $0.15$c along the line-of-sight and on the approaching side of the ejecta. Dashed white circles mark velocities of 0.1, 0.2, and 0.3c as in Fig.~\ref{fig:modelprop-0.5}.  }
    \label{fig:polmaps}
\end{figure*}

\section{Discussions}\label{sec:Discussions}

Overall, our knowledge of the properties of kilonova emission is limited. To improve our understanding, we performed 3-D MC radiative transfer simulations to predict the photometric and polarimetric behaviour for three different models, namely \texttt{Wind}, \texttt{Jet49}, and \texttt{Jet51}. From these simulations, we presented light curves for 11 different inclination angles from $0 \degr$ to $90\degr$ for $u,g,r,i,z,J,H,K$ filters. In addition, we also presented spectropolarimetric results along with polarization curves. We have made improvements in previous models presented in \citet{Bulla_2019, Bulla_2021} by including all the main components expected in binary neutron star mergers. In addition, the models are updated to include the time evolution of ejecta properties such as opacity, density, temperature, and they include time- and electron fraction-dependent nuclear heating rates \citep{Rosswog2022}.

Our models build on the setup from \citet{Nativi_2021} for the case of the lanthanide-poor disc with the addition of an extra component of lanthanide-rich dynamical ejecta and improved opacities. Our light curve behaviour is different from what was reported in \citet{Nativi_2021} (Fig. 4 in their paper). The light curves from our simulations have higher apparent magnitude compared to the previous results in \citet{Nativi_2021}. The values presented in this paper are closer to the apparent magnitude observed for GW 170817 KN AT2017gfo \citep{Andreoni_2017,Arcavi_2017,Chornock_2017,Cowperthwaite_2017,Drout_2017, Evans_2017, Kasliwal_2017,Tanvir_2017,Pian_2017,Troja_2017,Smartt_2017, Utsumi_2017,Valenti_2017}. In both cases, the variation with inclination angles is greatest for the \texttt{Wind} model. However, the overall variation with viewing angle is much higher in our results. This behaviour can be attributed to the variation in ejecta density distribution. Due to the presence of dynamical ejecta with higher density, polar viewing angles are always brighter than viewing angles closer to the equator. 

\begin{figure}
    \centering
\includegraphics[scale=0.36]{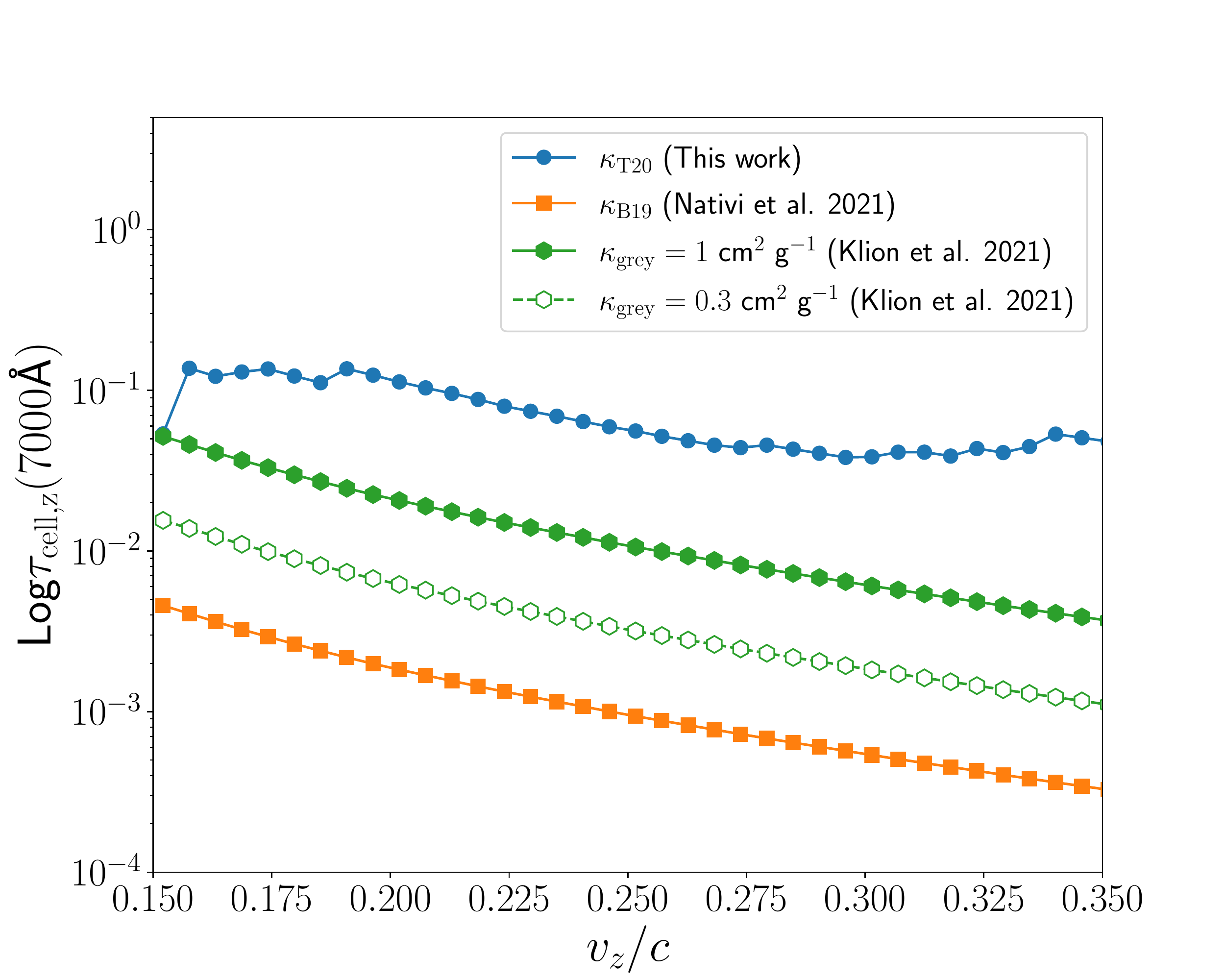}
    \caption{Optical depth within a cell as a function of velocity along the $z$ (jet) axis. Curves are shown for the \texttt{Jet51} model and for different assumptions for the opacities $\kappa$: state-of-the-art opacities from \citet{Tanaka_2020} adopted in this work ($\kappa_{\rm T20}$, blue circles), analytic and uniform opacities from \citet{Bulla_2019b} adopted in \citet{Nativi_2021} ($\kappa_{\rm B19}$, orange squares), grey opacities with $\kappa_{\rm grey}=1$\,cm$^2$\,g$^{-1}$ adopted in \citet{Klion_2021} (filled green hexagons) and grey opacities with $\kappa_{\rm grey}=0.3$\,cm$^2$\,g$^{-1}$ (open green hexagons) found to reproduce well detailed iron-group-like opacities used in \citet[][see their figure 9]{Klion_2021}. Optical depths are calculated within each cell at 1\,d after the merger. For the two sets of non-grey opacities, optical depths are calculated at $7000$\AA{} as $\tau_{\rm cell}=(\kappa_{\rm es}+\kappa_{\rm bb})\,\rho\,dr$, where $\kappa_{\rm es}$ and $\kappa_{\rm bb}$ are the Thomson scattering and bound-bound opacities and $dr$ is the cell width.}
    \label{fig:cf_nativi_klion}
\end{figure}

In contrast to previous studies \citep{Nativi_2021, Klion_2021}, we do not see the presence of a jet making the light curve brighter near the polar viewing angle. On the contrary, we find that the presence of a jet leads to a decrease in brightness for orientations close to the jet axis and an increase in brightness for more equatorial viewing angles. As shown in Fig.~\ref{fig:cf_nativi_klion} for the \texttt{Jet51} case, this effect is due to the difference in opacity setup for models in this paper. For the opacities, \citet{Nativi_2021} used simple analytic functions and a bimodal uniform distribution (lanthanide-poor + lanthanide-rich depending on whether $Y_e$ was larger or smaller than 0.25, respectively, \citealt{Bulla_2019b}), leading to ejecta that were optically thin in the regions where material from the wind are spread out by the jet ($\gtrsim0.15$c, orange squares). In contrast, the state-of-the-art opacities from \citet{Tanaka_2020} used in this work depend on local properties of the ejecta ($\rho$, $T$ and $Y_e$) and lead to moderately optically-thick regions for a wide range of velocity above $\gtrsim0.15$c (cyan squares). The integrated optical depth from $0.15$c to the grid boundary at $1$\,d after the merger is equal to $\sim$ 2.68 in our work while $\sim0.05$ when using the opacities in \citet{Nativi_2021}. Hence, in our work, photons that are emitted towards polar viewing angles can get scattered by material redistributed by the jet close to the jet axis and re-emitted with longer wavelengths towards the equatorial viewing angles, decreasing (increasing) the kilonova brightness for face-on (intermediate/edge-on) view of the system. Although we do not have information about optical depths in the models by \citet{Klion_2021}, we note that adopting in our models their grey opacities $\kappa_{\rm grey}=1$ and $\kappa_{\rm grey}=0.3$\,cm$^2$\,g$^{-1}$ (found to reproduce well their simulation with iron-group-like opacities) would lead to optically thin ejecta along the jet axis as in \citet{Nativi_2021} (filled and open green hexagons, respectively).  Although the opacities that we used in this work  \citep{Tanaka_2020} are more reliable than those employed in \citet{Nativi_2021} and \citet{Klion_2021}, it is worth stressing
that large uncertainties in $r-$process opacities remain, and it is difficult to quantify by how much our light curve results may be impacted by them.

This work also examined the impact of jets and dynamical ejecta on polarization signals. We present spectropolarimetry and polarization curves for the three different models. One of the major differences compared to previous results from \citet{Bulla_2019} and \citet{Bulla_2021} is that the polarization signal is present even at relatively late times like 2 to 3 days after the merger. This can be attributed to the difference in the opacity implementation in the simulations presented in this paper compared to the previous models. In addition, we observe that the presence of the jet component increases the overall polarization signal and the energy of the jet component also has an impact. Model \texttt{Jet51} shows the highest level of polarization out of the three models in $rizJ$ filters. This could be due to a further break in symmetry in these models due to the presence of the jet. In Fig.~\ref{fig:modelprop-0.5}, we can see scattering opacity present in the jet direction which is more prominent in \texttt{Jet51} case. This can scatter and polarize the photons which add to the detected polarization signal. We see the highest levels of polarization from intermediate angles of $26\degr$ and $66 \degr$. 

For model \texttt{Wind}, we overall do not see a significant polarization signal for any epoch. In the case of model \texttt{Jet49}, we detect negative Stokes $q$ for shorter wavelengths at 1.3 and 2.4 days after the merger. However, for 3.8 and 4.2 days after the merger, the polarization signal is positive Stokes $q$ at wavelengths between $1.0$ and $1.75 \micron$. For model \texttt{Jet51}, the sign of stokes $q$ depends on the viewing angles. For $26 \degr$, we observe positive Stokes $q$ for all the epochs and wavelengths, however for other angles the sign of $q$ flips from some negative to overall positive at days after 3.8. 
Hence, observation of significant polarization from kilonova can help us differentiate between the structure with or without the jet, and also the sign of observed Stokes $q$ can help with constraining the energetics of the jet.

As mentioned in Section~\ref{sec:Methods}, all three models assume a lanthanide-poor composition for the ``secular'' ejecta since a lanthanide-rich composition is expected to give no polarization. Polarimetric observations of future kilonovae will help constrain the composition of the ejecta and provide a smoking gun for the presence (detection) or absence (non-detection) of a lanthanide-poor component \citep{Bulla_2019,Bulla_2021}.

Polarization curves along with light curves can constrain the inclination angle of the system. If the density distribution is constrained via light curves then we can use polarization measurements to constrain the inclination angle. Depending on the model, the detectable polarization signal depends on the viewing angle and time since the merger. Thus, our models show that the combined observations of polarimetry and photometry will be powerful in our understanding of the kilonova ejecta structure.

\subsection{AT2017gfo }
One of the most well-studied kilonovae so far is the one associated with the gravitational-wave event GW170817, namely AT2017gfo \citep{Coulte_2017}. There are extensive observations data on this event, from spectra and light curves to polarization observations. In Fig.~\ref{fig:lc}, we have overplotted AT2017gfo observations data as the open circles for all the filters. We find that our models predict higher apparent magnitudes compared to the previous model by \citet{Nativi_2021}. This higher apparent magnitude are closer to the observational data from AT2017gfo \citep{Andreoni_2017, Arcavi_2017, Chornock_2017, Cowperthwaite_2017, Drout_2017, Evans_2017, Kasliwal_2017, Tanvir_2017, Pian_2017, Troja_2017, Smartt_2017, Utsumi_2017, Valenti_2017}. Qualitatively, we find a similar general trend between the observations and the models. However, we do not find one particular inclination angle to match the observational data. From the observed superluminal motion of the jet in radio images, \cite{Mooley2022} constrained the inclination angle to $\sim 19-25 \degr$ for AT2017gfo. We find that the models with \texttt{Jet} match better with the observed values at an inclination angle closer to $25 \degr$ ($\cos\theta_{\rm obs}=0.9$). However, we note that these simulations were not performed for AT2017gfo but were more generalized scenarios.

\citet{Covino_2017} reported a low level of linear optical polarization signal for AT2017gfo. With $5\sigma$ significance, they observed a polarization value of $0.5 \pm 0.7 \%$ at 1.46 days after the merger. They attribute this signal to interstellar polarization induced by galactic dust and the kilonova being intrinsically unpolarized. This agrees well with our polarization predictions from the models as shown via the polarization curve in Fig.~\ref{fig:polcurve}. At 1.46 days after the merger, our models predict polarization less than $0.2 \%$ for viewing angles $\lesssim25 \degr$. In addition, recently \citet{Sneppen_2023} showed that AT2017gfo was a highly symmetric explosion at early times using the Sr+ P Cygni profile along with kilonova blackbody features. Spherical ejecta would be consistent with the small polarization level observed in AT2017gfo, but potentially in conflict with numerical-relativity simulations \citep[see][for a review]{Nakkar_2020} and kilonova modeling suggesting the presence of at least two ejecta components with different geometries and compositions \citep[e.g.][]{Perego2017,Kawaguchi2018,possis2}.

 \section{Conclusions} \label{sec:Conclusions}
We have presented results for the state-of-the-art 3-D MC radiative transfer simulations using \textsc{possis}. These simulations are an improvement to the previous models due to the realistic merger ejecta distribution and opacities for this ejecta and the evolution of these quantities with time. We present predictions of both photometric and polarimetric behaviour of the emission from binary neutron star mergers for the cases \texttt{Wind}, \texttt{Jet49}, and \texttt{Jet51}. From our simulations, we concluded:
\begin{itemize}
    \item \textbf{Light curves} from these simulations show that color evolution is highly dependent on the presence of a jet. Thus, observations in a few different filters like $u,g,z$ can differentiate among these models. 
    \item \textbf{Polarimetric} results show that the presence of a jet has some impact on the detected level of polarization. We see jets with higher energy produce a higher level of polarization. 
    
    \item Observing polarization evolution with time in a few filters such as $rizJ$ filters can help us differentiate among different density structures.  
    
\end{itemize}

We note however that the predicted level of polarization is generally low. This is likely model-dependent given our choice to focus on specific realizations of binary neutron star mergers. In contrast, the increase in polarization from the presence of a jet is likely more robust due to the expectation that the jet will spread out electron-scattering plumes to wider regions.

The analysis presented in this work demonstrates how the combination of light curves and polarimetric observations of kilonova in a few different filters can provide valuable information about the event such as its inclination angle, and the presence of a jet in the ejecta. The predicted low levels of polarization make observations with current polarimeters a challenge for a new KN in the near future. However, with future,  more sensitive polarimeters, this level of polarization can be observable. Currently, observers can utilize these simulations to prepare for the best observing strategies for future kilonovae.

\section*{Acknowledgements}
The simulations were performed on resources provided by the Swedish National Infrastructure for Computing (SNIC) at Kebnekaise partially funded by the Swedish Research Council through grant agreement no. 2018-05973, as well as on the GCS Supercomputer SuperMUC-NG at the Leibniz Supercomputing Centre (LRZ) [project pn29ba].
MS was supported by an STFC consolidated grant number (ST/R000484/1) to LJMU. This work was supported by the European Union’s Horizon 2020 Programme under the AHEAD2020 project (grant agreement n. 871158).SR has been supported by the Swedish Research Council (VR) under 
grant number 2020-05044, by the research environment grant
``Gravitational Radiation and Electromagnetic Astrophysical
Transients'' (GREAT) funded by the Swedish Research Council (VR) 
under Dnr 2016-06012, by the Knut and Alice Wallenberg Foundation
under grant Dnr. KAW 2019.0112,   by the Deutsche 
Forschungsgemeinschaft (DFG, German Research Foundation) under 
Germany’s Excellence Strategy – EXC 2121 ``Quantum Universe'' 
– 390833306 and by the European Research Council (ERC) Advanced 
Grant INSPIRATION under the European Union’s Horizon 2020 research 
and innovation program (Grant agreement No. 101053985).

\section*{Data Availability}
The light curves for the models used in this study will be made available at \url{https://github.com/mbulla/kilonova_models}. The \textsc{possis} code used to simulate the light curves is not publicly available.



\bibliographystyle{mnras}
\bibliography{kilonova} 







\bsp	
\label{lastpage}
\end{document}